\documentclass[twocolumn,english,aps,pra,reprint,superscriptaddress,showpacs,showkeys,longbibliography]{revtex4-1}
\usepackage[english]{babel}
\usepackage{amsmath,amssymb,bbm,mathrsfs,bm,braket,color,graphicx,comment}
\usepackage[colorlinks,citecolor=blue,urlcolor=blue]{hyperref}
\usepackage[mathscr]{euscript}

\newcommand{\Z}{\mathbb{Z}}
\newcommand{\sx}{\sigma^x}

\newcommand{\sz}{\sigma^z}
\newcommand{\id}{\mathbbm{1}}
\newcommand{\heff}{H_{\mathrm{eff}}}
\newcommand{\usw}{U_{\mathrm{SW}}}

\newcommand{\abs}[1]{\left\lvert #1 \right\rvert}

\begin{document}

\title{Programmable superpositions of Ising configurations}

\author{Lukas M. Sieberer}

\affiliation{Department of Physics, University of California, Berkeley,
  California 94720, USA}

\author{Wolfgang Lechner}

\email{w.lechner@uibk.ac.at}
\affiliation{Institute for Theoretical Physics, University of Innsbruck, A-6020
Innsbruck, Austria}
\affiliation{Institute for Quantum Optics and Quantum Information of the Austrian
Academy of Sciences, A-6020 Innsbruck, Austria}

\date{\today}
\begin{abstract}
  We present a framework to prepare superpositions of bit strings, i.e.,
  many-body spin configurations, with deterministic programmable
  probabilities. The spin configurations are encoded in the degenerate ground
  states of the lattice-gauge representation of an all-to-all connected Ising
  spin glass. The ground state manifold is invariant under variations of the
  gauge degrees of freedom, which take the form of four-body parity constraints.
  Our framework makes use of these degrees of freedom by individually tuning
  them to dynamically prepare programmable superpositions. The dynamics combines
  an adiabatic protocol with controlled diabatic transitions. We derive an
  effective model that allows one to determine the control parameters
  efficiently even for large system sizes.
\end{abstract}




\maketitle

\section{Introduction}
\label{sec:introduction}

The realization of quantum many-body superpositions is a cornerstone of current
developments in quantum simulation
experiments~\cite{cirac2012goals,RevBloch,RevNori,Blatt2007,Zeiher2016,Schoelkopf2010,Lukin2017,Brune}. The
aim is to achieve deterministic tunability over individual constituents of the
quantum state in an experiment. Superpositions of bit strings (encoded in spin
configurations) have been recently proposed as a key to quantum machine learning
applications~\cite{qram,qramarchitecture,lloydml}. The challenge is thus to
prepare a superposition of a polynomial number $M$ bit strings in the
$2^N$-dimensional state space of $N$ qubits. From an experimental point of view,
the technique of adiabatic state preparation of spin
models~\cite{Nishimori1998,Farhi2000,amin2015,Boixo2016}, which recently gained
considerable interest as a tool to solve optimization problems, might serve as
an effective method to prepare such states. If the ground state of the final
Hamiltonian in an adiabatic protocol is energetically degenerate, the final
state of the protocol is a superposition of the configurations in the degenerate
manifold~\cite{matsuda2009ground,Boixo2014}. However, the amplitudes of this
state are governed by the details of the dynamics and the populations can be
exponentially biased~\cite{KatzgraberFair}, leaving open the challenge to
deterministically program these probabilities.

In this paper, we introduce a framework to generate superpositions of a
polynomial number of bit strings with programmable squared amplitudes via an
adiabatic-diabatic state preparation. The $M$ different bit strings of length
$N$ are encoded in the ground-state manifold of the lattice-gauge representation
of an all-to-all connected Ising model with $K = N (N - 1)/2$
spins~\cite{lechner2015}. Thus, a polynomial number $M$ of configurations
represent the ground-state manifold in an exponential $2^N$-dimensional space.

We show that the gauge degrees of freedom in this representation, i.e., the
constraints, allow one to shape the quantum dynamics and program the final
amplitudes of these $M$ configurations. We develop an effective $M$-dimensional
theory. In the reduced Hilbert space of the effective model, the parameters can
be determined efficiently, even for system sizes that cannot be solved on
current classical computers (i.e., more than 50 qubits).
\begin{figure}
  \centering
  \includegraphics[width=6.5cm]{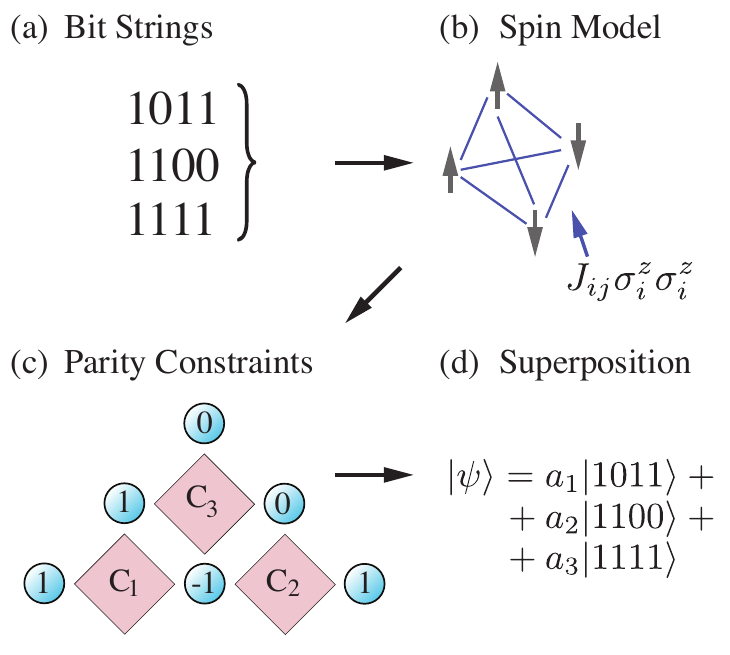}
  \caption{Schematic outline of the protocol to store classical bit strings in a
    quantum superposition with programmable probabilities $\abs{a_n}^2$. (a) The
    input are $M$ classical bit strings. (b) An all-to-all connected classical
    spin model with Hamiltonian $H = \sum_{i,j} J_{ij} \sz_i \sz_j$ is
    constructed such that the bit strings are its degenerate ground states. (c)
    The spin model is translated to a lattice gauge model with qubits (blue
    spheres) and with four-body constraint strengths $C_p$ (red
    squares)~\cite{lechner2015}. (d) The adiabatic passage of this model yields
    a superposition of the $M$ degenerate ground states. The probabilities can
    be tuned on demand by choosing the values of the $C_p$ appropriately.}
  \label{fig:fig1}
\end{figure}

The remainder of this paper is organized as follows: We begin in
Sec.~\ref{sec:protocol} by describing our state preparation
protocol. Section~\ref{sec:effective-model} is devoted to a detailed account of
the key step of the protocol, i.e., the adiabatic-diabatic dynamics leading to
the desired superposition state, and its description in terms of an effective
model. An example that illustrates the method is presented in
Sec.~\ref{sec:example}. We conclude in Sec.~\ref{sec:conclusions} and provide an
outlook on future research directions.

\section{Protocol}
\label{sec:protocol}

The basis of the protocol is the lattice-gauge representation of an all-to-all
connected spin
model~\cite{lechner2015,lhzrydberg,Rocchetto2016,Puri2017,leib2016transmon},
\begin{multline}
  \label{eq:hlhz}
  H(t) = - A(t) \sum_{i = 1}^K \sx_i - B(t) \sum_{i = 1}^K J_i \sz_i \\ - C(t)
  \sum_{p = 1}^{K-N+1} C_p \sz_{n_p} \sz_{w_p} \sz_{s_p} \sz_{e_p}.
\end{multline}
Here, $\sigma^{x,y,z}_i$ are the Pauli operators acting on spins which are
arranged on a two-dimensional square lattice with sites $i$, $K = N (N - 1)/2$,
and the sum in the last term runs over the $K - N + 1$ constraints. Each
constraint with index $p$ involves the spins in the north, west, south, and east
of the constraint site, as is indicated in Eq.~\eqref{eq:hlhz} with subscripts
$n_p, w_p, s_p,$ and $e_p$. The functions $A(t) = t/T$ and
$B(t) = C(t) = 1 - t/T$ are linear switching functions, and $T$ is the run time
of the protocol. Initially, $A(0) = 1$ and $B(0) = C(0) = 0$, while at the end
of the protocol $A(T) = 0$ and $B(T) = C(T) = 1$. We measure energies (and
inverse times) in units of the transverse field strength, which is thus set to
one. The qubits in Eq.~\eqref{eq:hlhz} represent connections between logical
spins of the all-to-all connected Ising model. In the mapping to the
lattice-gauge representation, additional degrees of freedom are introduced,
which are then removed by the constraints in the last term of
Eq.~\eqref{eq:hlhz} (see Ref.~\cite{lechner2015} for details). The constraint
strengths $C_p$ are gauge degrees of freedom. Tuning each of them individually
does not change the low-energy subspace in the final Hamiltonian in which the
bit strings are encoded. Thus the model features $(N - 2)(N - 3)/2$ additional
parameters as compared to the original spin-glass formulation. We use these
additional parameters to systematically design the quantum paths of an
adiabatic-diabatic protocol, which allows us to program the squared amplitudes
of the configurations in the final state.

Our protocol, which is illustrated in Fig.~\ref{fig:fig1}, aims at storing a
polynomial number $M$ of bit strings $x_n = 01011 \dots$ where
$n = 1, \dotsc, M$, as a quantum superposition state
$\ket{\psi} = a_1 \ket{x_1} + a_2 \ket{x_2} + \dotsb$ with programmable
probabilities $p_n = \abs{a_n}^2$. This superposition state thus encodes
classical data corresponding both to the bit strings $x_n$ and the corresponding
probabilities $p_n$. We regard the states $\ket{x_n} = \ket{01011 \dots}$ as
product states in the $\sz$ basis, with individual bits $x_{n,i} = 0, 1$
corresponding to eigenvalues $\pm 1$ of $\sz_i$.

The protocol consists of the following steps:
\begin{itemize}
\item[(i)] The bit strings are encoded as degenerate ground states of a
  classical spin model [see Fig.~\ref{fig:fig1}~(a) and~(b)]. This is achieved
  with a Hamiltonian of the from $H = \sum_{n=1}^M \ket{x_n} \bra{x_n}$ which
  can be approximated, for example, via plated solutions~\cite{PlantedSolutions}
  or Hopfield networks \cite{hopfield1982neural} in the form of an all-to-all
  connected model with energy $\sum_{i,j} J_{ij} \sz_i \sz_j$. The details of
  different strategies of this classical encoding will be discussed
  elsewhere~\cite{tobepublished}. We note that in this step a polynomial number
  of configurations is selected out of a $2^N$-dimensional space.
\item[(ii)] This spin Hamiltonian is reformulated in the parity-constraint model
  introduced in Ref.~\cite{lechner2015} with Hamiltonian $H(t)$ in
  Eq.~\eqref{eq:hlhz} [see Fig.~\ref{fig:fig1}~(c)]. Logical bit strings $x_n$
  are thus translated to physical bit strings $z_n$, representing spin
  configurations in the lattice-gauge model.
\item[(iii)] The key step, and focus of this work, is the dynamics that leads to
  the desired superposition. The system is initialized in the trivial and
  non-degenerate ground state of $H(0)$, in which all spins are aligned by the
  transverse field. Evolution with $H(t)$ yields a superposition of all
  configurations in the ground state manifold of $H(T)$,
  $\ket{\psi} = \sum_{n=1}^M a_n \ket{z_n}$, where the probabilities
  $\abs{a_n}^2$ can be tuned on demand by adjusting the control parameters
  $C_p$. These parameters $C_p$ are determined from static properties of an
  effective Hamiltonian, and can be fine-tuned via an iterative protocol. Note
  that only a polynomial number of parameters is required as $M$ is assumed to
  be polynomial.
\end{itemize}

\section{Effective model}
\label{sec:effective-model}

The quantum dynamics during step (iii) of the protocol can be understood in
terms of a simple effective model which we derive in the following. In
Secs.~\ref{sec:adiab-diab-dynam} and~\ref{sec:deriv-effect-model}, we introduce
the adiabatic manifold and the effective Hamiltonian, based on the observation
that during a slow sweep in the time-dependent Hamiltonian~\eqref{eq:hlhz} the
evolution is restricted to a manifold of the lowest-lying instantaneous energy
eigenstates. The effective Hamiltonian governing the dynamics within this
manifold can be determined by a Schrieffer-Wolff transformation, which we
describe in Sec.~\ref{sec:calc-effect-hamilt}. Using the effective Hamiltonian
we determine the control parameters for the protocol in
Secs.~\ref{sec:estimation-t_d} and~\ref{sec:determ-contr-param}.

\subsection{Adiabatic-diabatic dynamics}
\label{sec:adiab-diab-dynam}

The quantum adiabatic algorithm with a single non-degenerate ground state
of the ``problem Hamiltonian'' $H(T)$ is successful if adiabaticity is
maintained throughout the time evolution.  Here, in contrast, we consider the
case of an $M$-fold degenerate ground state manifold of the problem
Hamiltonian~\cite{KatzgraberFair}. Thus, the gap closes as $t \to T$ and the
dynamics can never be fully adiabatic. Nevertheless, by making $T$ large,
transitions out of the \textit{adiabatic manifold} (AMF), which is spanned by
the $M$ lowest-lying instantaneous eigenstates
$\{ \ket{\varphi_n(t)} \mid n = 1, \dotsc, M \}$ of $H(t)$ and evolves for
$t \to T$ into the degenerate ground state manifold of $H(T)$, are
suppressed. Our protocol aims at controlling the \emph{diabatic dynamics within
  the AMF} by tuning the constraint strengths $C_p$. In this sense, the final
state is prepared \textit{adiabatic-diabatically}.

Within the AMF, the dynamics consists of two different regimes: up to a
characteristic time $t_d$ the dynamics is adiabatic and the system remains in
the instantaneous ground state $\ket{\varphi_1(t)}$; from $t_d$ to $T$ the
dynamics can be considered as a sudden quench of the Hamiltonian parameters,
during which the state of the system $\ket{\psi(t)}$ essentially remains frozen,
i.e., $\ket{\psi(T)} \approx \ket{\psi(t_d)} \approx \ket{\varphi_1(t_d)}$. (In
the last equality, we omitted dynamical and Berry phases since we are interested
solely in programming the \emph{probabilities} of the final superposition
state.) A detailed discussion of this
Kibble-Zurek-inspired~\cite{Kibble1980,*Kibble2001,*Zurek1985,*Zurek1993,*Zurek1996}
approximation in the context of the Landau-Zener problem is provided in
Ref.~\cite{Damski2005}. The \emph{dynamical} problem of controlling diabatic
transitions is thus (at least approximately) reduced to the \emph{static}
problem of choosing the constraint strengths $C_p$ to tune the composition of
the instantaneous ground state $\ket{\varphi_1(t_d)}$ at time $t_d$. Below, we
derive the effective Hamiltonian $\heff(t)$ in the AMF that allows us to do this
efficiently. The characteristic time scale $t_d$ can also be determined within
the effective model as described in Sec.~\ref{sec:estimation-t_d}.

\subsection{Derivation of the effective model}
\label{sec:deriv-effect-model}

To put what we have described so far on a more formal base, we decompose the
total Hilbert space as $\mathscr{H} = \mathscr{P}(t) \oplus \mathscr{Q}(t)$,
where $\mathscr{P}(t)$ is the AMF that is spanned by the $M$ lowest-lying
instantaneous eigenstates, $\ket{\varphi_n(t)}$ with $n = 1, \dotsc, M$, of the
lattice-gauge Hamiltonian $H(t)$~\eqref{eq:hlhz}. These states become degenerate
for $t \to T$, and consequently the dynamics within $\mathscr{P}(t)$ becomes
diabatic, whereas a finite gap between the energies of the states in
$\mathscr{P}(t)$ and higher-lying excited states spanning the subspace
$\mathscr{Q}(t)$ is maintained throughout the protocol. Hence, transitions out
of the AMF $\mathscr{P}(t)$ can be suppressed effectively by choosing the run
time $T$ of the protocol large enough. Strictly speaking, states within
$\mathscr{P}(t)$ are also degenerate with states of $\mathscr{Q}(t)$ at $t = 0$.
However, the system is initialized in the instantaneous ground state which is
non-degenerate at $t = 0$ and separated from all excited states by a finite gap,
so there are no diabatic transitions at short times. A state
$\ket{\psi(t)} \in \mathscr{P}(t)$ can be written as
\begin{equation}
  \label{eq:psi_adiabatic_basis}
  \ket{\psi(t)} = \sum_{n = 1}^M \alpha_n(t) \ket{\varphi_n(t)},
\end{equation}
and the Schr\"odinger equation, projected to $\mathscr{P}(t)$, takes the form
\begin{equation}
  \label{eq:SE_adiabatic_basis}
  \begin{split}
    i \dot{\alpha}_n(t) & = \sum_{n' = 1}^M A_{n n'}(t) \alpha_{n'}(t), \\ A_{n
      n'}(t) & = \braket{\varphi_n(t) | \left( H(t) - i \frac{d}{dt} \right) |
      \varphi_{n'}(t)}.
  \end{split}
\end{equation}
The dynamics within the AMF is thus governed by the matrix elements $A_{nn'}(t)$
which are in turn determined by the instantaneous eigenstates
$\ket{\varphi_n(t)}$. Finding these states by diagonalizing $H(t)$ can be
achieved in a two-step procedure: (i) The space $\mathscr{P}(t)$ of the $M$
lowest-lying states is decoupled from excited state space $\mathscr{Q}(t)$ by a
Schrieffer-Wolff (SW) transformation $\usw(t)$~\cite{Schrieffer1966}. As we
describe in detail in Sec.~\ref{sec:calc-effect-hamilt}, the SW transformation
can be obtained as a perturbative expansion in the transverse field. (ii) The
resulting effective Hamiltonian $\heff(t) = \usw(t) H(t) \usw^{\dagger}(t)$ is
diagonalized within the reduced space by applying another unitary transformation
$U_0(t)$. Hence, the instantaneous eigenstates of $H(t)$ that span
$\mathscr{P}(t)$ can be written as
\begin{equation}
  \label{eq:SW_eigenstate}
  \ket{\varphi_n(t)} = \usw^{\dagger}(t) U_0^{\dagger}(t) \ket{z_n}.
\end{equation}
Inserting this representation in Eq.~\eqref{eq:SE_adiabatic_basis} we obtain
\begin{multline}
  \label{eq:A}
  A_{nn'}(t) = \bra{z_n} \left[ U_0(t) \left( \heff(t) - i \usw(t)
      \dot{U}_{\mathrm{SW}}(t)^{\dagger} \right) \right. \\ \left. \times
    U_0(t)^{\dagger} - i U_0(t) \dot{U}_0(t)^{\dagger} \right] \ket{z_{n'}}.
\end{multline}
While the perturbative expansion of $\usw(t)$ can be derived without specifying
the exact superposition state that should be prepared [which is encoded in the
local field term in Eq.~\eqref{eq:hlhz}], the unitary transformation $U_0(t)$
explicitly depends on these details. For this reason it is more convenient to
work in a basis in which $\heff$ is not diagonal, i.e.,
\begin{equation}
  \label{eq:beta_alpha}
  \beta_n(t) = \sum_{n' = 1}^M \braket{z_n | U_0^{\dagger}(t) | z_{n'}} \alpha_{n'}(t).
\end{equation}
Then, the Schr\"odinger equation~\eqref{eq:SE_adiabatic_basis} takes the form
\begin{equation}
  \label{eq:SE_beta}
  i \dot{\beta}_n(t) = \sum_{n'} B_{n n'}(t) \beta_{n'}(t),
\end{equation}
where
\begin{equation}  
  B_{n n'}(t) =
  \braket{z_n | \left( \heff(t) - i \usw(t) \dot{U}_{\mathrm{SW}}^{\dagger}(t)
    \right) | z_{n'}}.
\end{equation}
Due to the derivative with respect to time, the second term in this matrix
element is suppressed by an additional factor of $1/T$ (recall that $T$ is the
dimensionless run time of the protocol measured in units of the inverse local
field strength), and can thus be dropped~\footnote{As described in
  Sec.~\ref{sec:calc-effect-hamilt}, both $\heff(t)$ and $\usw(t)$ can be
  expressed as perturbative expansions in $1 - t/T$; then, for each term of
  order $\left( 1 - t/T \right)^n$ in $\heff(t)$, the corresponding term in
  $\usw(t) \dot{U}_{\mathrm{SW}}^{\dagger}(t)$ is of order
  $\left( 1/T \right) \left( 1 - t/T \right)^{n - 1}$. To determine the control
  parameters for the protocol, we evaluate the effective Hamiltonian at $t_d$,
  which is typically a finite fraction of $T$. Then, $1/T \ll 1 - t_d/T$, and
  the second term in $B_{n n'}(t)$ can safely be neglected.}. Then, in its
simplest form, the effective model is given by the equation of motion in
Eq.~\eqref{eq:beta_alpha} with
\begin{equation}
  \label{eq:B}
  B_{n n'}(t) \approx \braket{z_n | \heff(t) | z_{n'}}.
\end{equation}
We note, however, that including the second term in $B_{nn'}(t)$ to make the
effective model more accurate is straightforward. Having specified the dynamics
within the AMF and its description in terms of an effective Hamiltonian, we
proceed to discuss the perturbative expansion of the SW transformation
$\usw(t)$.

\subsection{Calculation of the effective Hamiltonian}
\label{sec:calc-effect-hamilt}

The SW transformation $\usw(t)$ decouples the AMF $\mathscr{P}(t)$ from the
manifold of excited states $\mathscr{Q}(t)$. To find $\usw(t)$, we can treat the
transverse field in the time-dependent Hamiltonian~\eqref{eq:hlhz} as a
perturbation. Clearly, such an expansion could not be justified at early times
of the evolution when $A(t) \approx 1$ while $B(t), C(t) \ll 1$ and the
transverse field is the dominant part of the
Hamiltonian~\eqref{eq:hlhz}. However, as explained in
Sec.~\ref{sec:adiab-diab-dynam}, we require our effective model to be accurate
only at $t_d$ when the dynamics becomes diabatic. This is the case towards the
end of the protocol when the strength of the transverse field $A(t_d)$ becomes
small and the gaps between the states $\ket{\varphi_n(t)}$ spanning the AMF
close. Moreover, below we will see that each order in perturbation theory is
accompanied by an additional factor of $1/C_p$ for one of the constraints
$p = 1, \dotsc, K - N + 1$ in Eq.~\eqref{eq:hlhz}, i.e., the effective expansion
parameter is $A(t_d)/C_p$ --- which is small by construction of the
lattice-gauge formulation~\cite{lechner2015}. We emphasize, however, that we use
perturbation theory only to construct $\usw(t)$ and thus solve the \emph{static}
problem of finding the instantaneous eigenstates of $H(t)$, whereas the
\emph{dynamics} is fundamentally non-perturbative (in the sense that it cannot
be described by, e.g., adiabatic perturbation theory~\cite{DeGrandi2010}).

Let us now consider the structure of the ensuing perturbative expansion. The
unperturbed Hamiltonian $H_0(t)$ comprises the local fields and the four-body
constraints in Eq.~\eqref{eq:hlhz},
\begin{equation}
  \label{eq:H0}
  H_0(t) = \frac{t}{T} \left( H_J + H_C \right),  
\end{equation}
where
\begin{equation}
  H_J = - \sum_{i = 1}^K J_i \sz_i, \quad H_C = - \sum_{p = 1}^{K-N+1} C_p
  \sz_{n_p} \sz_{w_p} \sz_{s_p} \sz_{e_p},
\end{equation}
while the perturbation is given by the transverse field,
\begin{equation}
  \label{eq:V}
  V(t) = \left( 1 - \frac{t}{T} \right) \sum_{i = 1}^K \sx_i,
\end{equation}
and the total Hamiltonian is thus
\begin{equation}
  \label{eq:H}
  H(t) = H_0(t) + V(t).
\end{equation}
$H_0(t)$ is diagonal in a basis of $\sz$-product states. In particular, the
states $\ket{z_n}$ where $n = 1, \dotsc, M$ span the degenerate ground state
manifold $\mathscr{P}_0$ of $H_0(t)$. The ground state manifold $\mathscr{P}_0$
coincides with the AMF at the end of the protocol when $t = T$, i.e.,
$\mathscr{P}(T) = \mathscr{P}_0$ and thus $\usw(T) = \id$, while at any $t < T$,
the SW transformation is a direct rotation between $\mathscr{P}(t)$ and
$\mathscr{P}_0$~\cite{Bravyi2011}. Then, the effective Hamiltonian defined by
$\heff(t) = \usw(t) H(t) \usw^{\dagger}(t)$ is block-diagonal when written in
the basis of eigenstates of $H_0(t)$ --- which, as pointed out above, is just
the basis of $\sz$-product states. In other words, the SW transformation
$\usw(t)$ decouples the space $\mathscr{P}(t)$ from the space of excited states
$\mathscr{Q}(t)$ by incorporating the effect of virtual transitions to
$\mathscr{Q}(t)$ in the effective Hamiltonian $\heff(t)$.

The form of the perturbative expansion of $\heff(t)$ is particularly transparent
since the perturbation $V(t)$, which is the sum of terms $\sx_i$, has the effect
of flipping single physical spins. In other words, applying the perturbation
once to a state $\ket{z_n}$ yields a superposition of states at a Hamming
distance of one, where the Hamming distance is measured with respect to the
encoded bit string $z_n$. The transverse field $V(t)$ thus (i) modifies the
energies (i.e., the diagonal elements of the effective Hamiltonian) of the
states $\ket{z_n}$ at second order in perturbation theory and (ii) couples
states $\ket{z_n}$ and $\ket{z_{n'}}$ (i.e., it generates off-diagonal elements
of $\heff$) at the order of their Hamming distance $h_{nn'}$. The second effect
(ii) can be understood by noting that applying the perturbation $V$ once to the
state $\ket{z_n}$ flips a single spin and therefore it has to be applied
$h_{nn'}$ times to connect the states $\ket{z_n}$ and $\ket{z_{n'}}$. We further
note, that since the qubits of the lattice-gauge representation encode the
relative orientation of spins in the original all-to-all spin glass
model~\cite{lechner2015}, flipping a single spin of a state in $\mathscr{P}$
always yields a state in $\mathscr{Q}_0 = \mathscr{Q}(T)$ (thus leading to an
additional factor of $1/C_p$ in the perturbative expansion as asserted
above). For this reason, Hamming distances between ground states are at least
two, and hence the lowest non-trivial order in perturbation theory that
contributes to $\heff(t)$ is two. From this discussion, we can already infer
that the matrix elements
$H_{\mathrm{eff}, n n'}(t) = \braket{n | \heff(t) | n'}$ of the effective
Hamiltonian take the following form to leading order in perturbation theory:
\begin{equation}
  \label{eq:heff-matrix-elements}
  \begin{split}
    H_{\mathrm{eff}, nn}(t) & = \frac{t}{T} e_0 + \frac{T}{t} \left( 1 -
      \frac{t}{T} \right)^2 e_n, \\ H_{\mathrm{eff}, nn'}(t) & = \left(
      \frac{T}{t} \right)^{h_{nn'} - 1} \left( 1 - \frac{t}{T} \right)^{h_{nn'}}
    g_{nn'},
  \end{split}
\end{equation}
Here, $t e_0/T$ is the ground-state energy of $H_0(t)$, which according to
Eq.~\eqref{eq:H0} depends linearly on time. The time dependence of the other
terms reflects the general structure of perturbation theory: For each order of
the perturbative expansion is a factor of $1 - t/T$, i.e., the strength of the
transverse field at time $t$, while the factors $T/t$ stem from the energy
denominators, which are differences of eigenenergies of the unperturbed
Hamiltonian $H_0(t)$. We note that the divergence of the factors $T/t$ for
$t \to 0$ need not bother us since as discussed above we are interested
primarily in the effective Hamiltonian at $t = t_d$. Below we describe how the
coefficients $e_n$ and $g_{nn'}$ can be obtained explicitly. They are given in
Eqs.~\eqref{eq:e},~\eqref{eq:g1323}, and~\eqref{eq:g12} for a specific
example. For Hamming distances $h_{nn'} > 2$, we note that the off-diagonal
elements in Eq.~\eqref{eq:heff-matrix-elements} vanish faster
$\sim \left( 1 - t/T \right)^{h_{nn'}}$ than the diagonal ones
$\sim \left( 1 - t/T \right)^2$ as $t \to T$. Therefore, the instantaneous
eigenstates $\ket{\varphi_n(t)}$ converge to the states $\ket{z_n}$ --- and, in
particular, not to linear combinations of these states. This corroborates that
preparing a superposition of the states $\ket{z_n}$ dynamically relies crucially
on diabatic transitions.

After these preliminaries, let us explicitly specify the perturbative expansion,
i.e., the calculation of $e_n$ and $g_{nn'}$ in
Eq.~\eqref{eq:heff-matrix-elements}. We adopt the notation of
Ref.~\cite{Bravyi2011}, and all results for the SW we use in the following can
be found there. In what follows, we denote by
$P = \sum_{n = 1}^M \ket{z_n} \bra{z_n}$ the projector on the ground-state
manifold $\mathscr{P}_0$, and by $Q = \id - P$ the projector on the excited
state space $\mathscr{Q}_0$. These projectors commute with the unperturbed
Hamiltonian, $[P, H_0] = [Q, H_0] = 0$. The ground state energy is $E_0$, i.e.,
we have $H_0 \ket{z_n} = E_0 \ket{z_n}$. According to Eq.~\eqref{eq:H0}, $E_0$
depends linearly on time. However, for simplicity, we suppress the dependence on
time of both $H_0$ and $E_0$. Since the aim of the SW transformation is to
decouple $\mathscr{P}(t)$ and $\mathscr{Q}(t)$, i.e., to bring the Hamiltonian
to block-diagonal form, it is useful to introduce the following superoperators:
\begin{equation}
  \label{eq:OD}
  \mathscr{D}(X) = P X P + Q X Q, \quad \mathscr{O}(X) = P X Q + Q X P.
\end{equation}
An operator $X$ is block-diagonal (block-off-diagonal) iff $\mathscr{D}(X) = X$
($\mathscr{O}(X) = X$). Any operator can be decomposed into block-diagonal and
block-off-diagonal components. In particular, for the perturbation we have
\begin{equation}
  \label{eq:Vdod}
  V = V_{\mathrm{d}} + V_{\mathrm{od}}, \quad V_{\mathrm{d}} = \mathscr{D}(V),
  \quad V_{\mathrm{od}} = \mathscr{O}(V).
\end{equation}
Finally, for the present case in which all states in $\mathscr{P}_0$ are
degenerate, the superoperator $\mathscr{L}$ defined in Ref.~\cite{Bravyi2011}
takes the simple form
\begin{equation}
  \label{eq:L}
  \mathscr{L}(X) = P X \frac{Q}{E_0 - H_0} - \frac{Q}{E_0 - H_0} X P.
\end{equation}
Explicit expressions for $\heff(t)$ up to fourth order in perturbation theory
are given in Ref.~\cite{Bravyi2011}, and we repeat them here for
completeness. This order of perturbation theory is sufficient for the example
discussed below which involves ground states with Hamming distances three and
four. Higher orders --- that are required to treat systems with ground states
with larger Hamming distances --- can be obtained through a systematic iterative
procedure. The general form of the effective Hamiltonian to fourth order in
perturbation theory is as follows:
\begin{equation}
  \label{eq:Heff_full}
  \begin{aligned}
    \heff^{(0)} & = H_0 P, & \heff^{(2)} &= \frac{1}{2} P [S_1, V_{\mathrm{od}}] P, \\
    \heff^{(1)} & = P V P, &    
    \heff^{(3)} & = \frac{1}{2} P [V_{\mathrm{od}}, \mathscr{L}([V_{\mathrm{d}},
    S_1])] P,
  \end{aligned}
\end{equation}
and
\begin{multline}
  \heff^{(4)} = \frac{1}{2} P \left( \frac{1}{4} [S_1, [S_1, [S_1,
    V_{\mathrm{od}}] ] ] \right. \\ \left. \vphantom{\frac{1}{2}} -
    [V_{\mathrm{od}}, \mathscr{L} ([V_{\mathrm{d}}, \mathscr{L}([V_{\mathrm{d}},
    S_1])])] \right) P,    
\end{multline}
where $S_1 = \mathscr{L}(V_{\mathrm{od}})$ is the first-order term in the
perturbative expansion of the generator $S = \ln(\usw)$ of the Schrieffer-Wolff
transformation~\cite{Bravyi2011}. For Hamming distances three and four between
the degenerate ground states in the example considered below, there are some
further simplifications. To name an example, we have $P V P = 0$, while
$P V Q V P$ has only diagonal elements, and $P V Q V Q V P$ is purely
off-diagonal with non-vanishing elements between states with Hamming distance
three etc. Using these simplifications we find:
\begin{equation}
  \label{eq:Heff}
  \begin{aligned}
    \heff^{(0)} & = E_0 P, & \heff^{(2)} & = P V \frac{Q}{E_0 - H_0} V P, \\
    \heff^{(1)} & = 0, &    
    \heff^{(3)} & = P \left( V \frac{Q}{E_0 - H_0} \right)^2 V P,     
  \end{aligned}
\end{equation}
and
\begin{multline}  
  \heff^{(4)} = P \left( V \frac{Q}{E_0 - H_0} \right)^3 V P \\
     - \frac{1}{2} \left[ P V \left( \frac{Q}{E_0 - H_0} \right)^2 V P V
      \frac{Q}{E_0 - H_0} V P \right. \\ \left. + P V \frac{Q}{E_0 - H_0} V P V \left(
        \frac{Q}{E_0 - H_0} \right)^2 V P \right].
\end{multline}
We note that $\heff^{(0)}, \heff^{(2)}$, and the last two terms in $\heff^{(4)}$
are diagonal whereas $\heff^{(3)}$ and the first term in $\heff^{(4)}$ have
non-zero elements only away from the diagonal. In the following, we omit the
subleading diagonal contributions to $\heff$ stemming from $\heff^{(4)}$.  Then,
the matrix elements of the effective Hamiltonian,
\begin{equation}
  \label{eq:heff}
  \begin{split}    
    H_{\mathrm{eff}, n n} & = E_0 + \bra{z_n} V
    \frac{Q}{E_0 -  H_0} V \ket{z_n}, \\    
    H_{\mathrm{eff}, n n'} & = \bra{z_n} \left( V \frac{Q}{E_0 - H_0}
    \right)^{h_{n n'} - 1} V \ket{z_{n'}},
  \end{split}
\end{equation}
take the simple form reported in Eq.~\eqref{eq:heff-matrix-elements} above. In
particular, the coefficients $e_n$ and $g_{nn'}$ can be written as
\begin{equation}
  e_n = \frac{t}{T} \left( 1 - \frac{t}{T} \right)^{-2} \bra{z_n} V
  \frac{Q}{E_0 -  H_0} V \ket{z_n},  
\end{equation}
and
\begin{multline}
  g_{nn'} = \left( \frac{t}{T} \right)^{h_{nn'} - 1} \left( 1 - \frac{t}{T}
  \right)^{-h_{nn'}} \\ \times \bra{z_n} \left( V \frac{Q}{E_0 - H_0}
  \right)^{h_{n n'} - 1} V \ket{z_{n'}}.
\end{multline}
We note that from the expressions for $H_0(t)$ in Eq.~\eqref{eq:H0} and $V(t)$
in Eq.~\eqref{eq:V} it is evident that $e_n$ and $g_{nn'}$ do not depend on
time. However, the constraint strengths $C_p$ enter both $E_0$ and $H_0$ and
thus allow us to tune the matrix elements of the effective
Hamiltonian. Consequently, also the composition of the eigenstates of $\heff(t)$
can be adjusted and, following the logic outlined in
Sec.~\ref{sec:adiab-diab-dynam}, we use this freedom to ensure that the
instantaneous ground state at the time $t_d$, at which the dynamics becomes
diabatic, equals the superposition state we seek to prepare.

Before we proceed to estimate the time scale $t_d$, we note that the effective
model, applied to the original all-to-all spin glass formulation, also provides
a general framework to address the problem of fair sampling of degenerate ground
states through quantum annealing~\cite{matsuda2009ground,KatzgraberFair}. In
particular, for spin-glass benchmark instances with controlled ground-state
degeneracy, the effective model can be used to obtain the output state of
quantum annealing with little computational effort. It is straightforward to
extend the effective model to study the impact of, e.g., more complex driving
Hamiltonians, on the composition of the output state. As we have shown here, in
the lattice-gauge formulation, fair sampling can be achieved due to the
additional ``tuning knobs'' provided by the parameters $C_p$.

\subsection{Estimation of $t_d$}
\label{sec:estimation-t_d}

To estimate the time $t_d$ at which the dynamics within the AMF becomes
diabatic, we treat the approach for $t \to T$ of each pair of levels
$E_n(t) - E_{n'}(t) \to 0$ where $n, n' = 1, \dotsc, M$ as an individual
Landau-Zener problem. The latter is characterized by a time-dependent velocity
and gap~\cite{LandauLifshitz3}:
\begin{equation}
  \label{eq:vDelta}
  v_{n n'} = \abs{\frac{d}{dt} \left( H_{\mathrm{eff}, n n} - H_{\mathrm{eff},
        n' n'} \right)}, \quad \Delta_{n n'} = H_{\mathrm{eff}, n n'},
\end{equation}
and $t_d$ is determined by the usual criterion that separates the diabatic from
the adiabatic regime in the Landau-Zener problem,
$v_{nn'}/\Delta_{nn'}^2 = \pi$. Of all $t_d$ found in this way for different
pairs of levels $n$ and $n'$, the smallest value indicates which transition
becomes diabatic first. In general, transitions with higher Hamming distances
have larger $t_d$. For the determination of the control parameters, i.e., the
constraint strengths, we use the smallest value.

\subsection{Determination of control parameters}
\label{sec:determ-contr-param}

Now we have all the tools at hand to determine the parameters $C_p$ that lead to
a final state $\ket{\psi(T)} = \sum_{n = 1}^M a_n \ket{z_n}$ with the required
probabilities $p_n = \abs{a_n}^2$. One has to calculate the ground state
$\ket{\varphi_1(t_d)} = \sum_{n = 1}^M b_n \ket{z_n}$ of $\heff(t_d)$ and find
the values $C_p$ that minimize the cost function
\begin{equation}
  \label{eq:cost_function}
  \Omega(\{b_n\}) = \sum_{n = 1}^M \left( \abs{b_n}^2 - p_n \right)^2.
\end{equation}
This gives the desired result since $\ket{\psi(T)} \approx \ket{\varphi_1(t_d)}$
and thus $a_n \approx b_n$ as explained above. For a given choice of the
probabilities $p_n$, typically the minimum of the cost function obtained in this
way is $\Omega(\{b_n\}) \approx 0$, which means that within our approximate
treatment the desired superposition could be prepared with fidelity close to
one. We investigate whether this is true for any set of probabilities
systematically for a specific example in Sec.~\ref{sec:ergod-solut-space}
below. To further improve the fidelity of the solution, one can iteratively
optimize the values of the $C_p$. In each iteration, the time evolution with
$\heff(t)$ is calculated to obtain the final state $\ket{\psi(T)}$ beyond the
sudden quench approximation, and the $C_p$ are updated to minimize the cost
function $\Omega(\{a_n\})$ evaluated for the final state~\footnote{Note, that
  the minimum $\Omega(\{a_n\}) = 0$ can only be reached if there are no
  transitions out of the AMF.}. As noted below
Eq.~\eqref{eq:heff-matrix-elements}, the perturbative expansion of $\heff(t)$
diverges for $t \to 0$. This, however, turns out to not be a severe obstacle in
practice: for the iterative optimization of the constraint strengths $C_p$, we
initialize the time evolution at a finite time $t_0 > 0$ in the instantaneous
ground state $\ket{\varphi_1(t_0)}$. In the examples we considered, we found
that the optimized $C_p$ are almost insensitive to the value of $t_0$ for
$0 < t_0 \lesssim t_d$. Since the optimization is carried out in the effective
$M$-dimensional model, it can be done efficiently even if the dimension of the
total physical Hilbert space $2^K$ is so large that the state preparation cannot
be simulated on a classical computer but still be performed on a quantum device.

\section{Example}
\label{sec:example}

Having described the protocol and the determination of the control parameters in
detail, let us now illustrate the method by the example shown in
Fig.~\ref{fig:fig1}. We take $M = 3$ bit strings $x_1 = 1011$, $x_2 = 1100$, and
$x_3 = 1111$, and for demonstration we prepare superpositions of the states
$\ket{x_n}$ with target probabilities $p_n = 1/M$ as well as $p_1 = 0.2$,
$p_2 = 0.3$, and $p_3 = 0.5$.

\subsection{Preparation of superposition states}
\label{sec:prep-superp-stat}

\begin{figure}
  \centering
  \includegraphics[width=8.6cm]{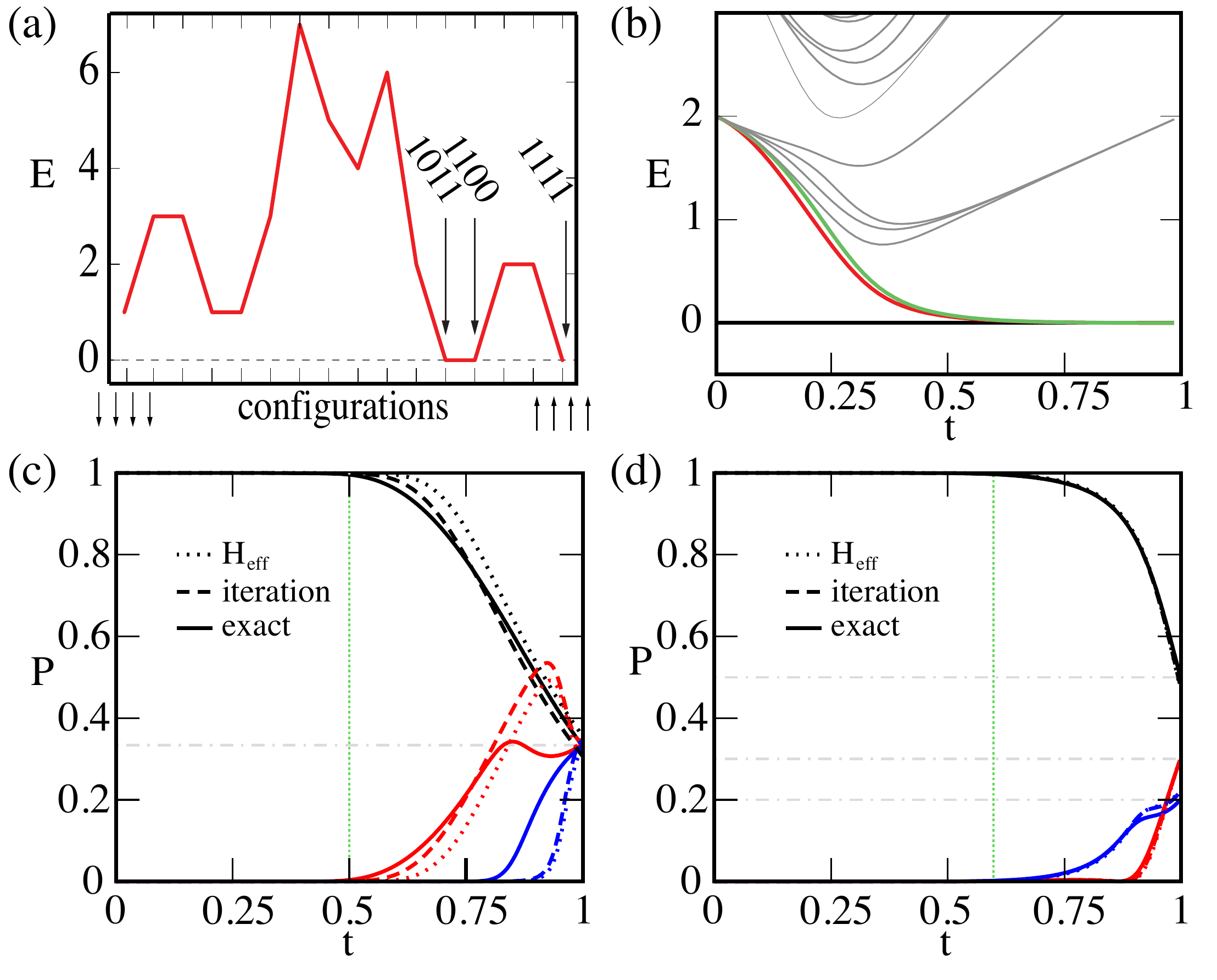}
  \caption{(a) Energy landscape of the classical spin model described in
    Sec.~\ref{sec:example}, which features global minima for configurations
    $1011$, $1100$ and $1111$. The symmetry of global inversion can be broken by
    an additional local field. (b) The energy spectrum of the time-dependent
    Hamiltonian Eq.~\eqref{eq:hlhz} shows a collapse to the degenerate states
    (black, red and green) of the final Hamiltonian. (c) Fair sampling:
    probabilities of the lowest $M$ instantaneous eigenstates of the
    time-dependent Hamiltonian Eq.~\eqref{eq:hlhz} for the example shown in
    Fig.~\ref{fig:fig1}. The time of diabaticity $t_d$ from the effective model
    is $t_d \approx 0.5 T$ (green dotted). The results from exact dynamics
    (solid), effective model (dotted) and improved iterative method (dashed) are
    all in good agreement. (d) Probabilities set to $p_1 = 0.2$, $p_2=0.3$, and
    $p_3=0.5$ with color code as in panel (c), where $t_d \approx 0.6 T$. In (c)
    and (d), time is measured in units of $T$.}
  \label{fig:results}
\end{figure}

In step (i) of the protocol described in Sec.~\ref{sec:protocol}, these bit
strings are encoded as the ground states of the Hamiltonian
$H = \sum_{i,j} J_{ij} \sz_i \sz_j$ with $J_{12} = J_{13} = J_{34} = 1$,
$J_{23} = -1$ and $J_{14} = J_{24} = 0$. The full energy landscape of $H$ is
shown in Fig.~\ref{fig:results}~(a), where the global $\Z_2$ symmetry is broken
by an additional local field of strength $h=1$. In step (ii) we switch to the
lattice-gauge representation~\cite{lechner2015} featuring $K = 6$ qubits and
three constraints as depicted in Fig.~\ref{fig:fig1}~(c). Due to the specific
arrangement of the qubits, the constraints with strengths $C_1$ and $C_2$ are
three-body interactions and $C_3$ is a four-body interaction (see
Appendix~\ref{sec:effect-hamilt-example} for
details). Figure~\ref{fig:results}~(b) shows the time-dependent spectrum of the
Hamiltonian~\eqref{eq:hlhz} with $C_1 = C_2 = C_3 = 4$. To determine the values
we have to assign to these parameters in order to prepare specific superposition
states in step (iii) of the protocol, we have to determine the effective
Hamiltonian $\heff(t)$. For this example, the matrix elements of $\heff(t)$ are
given by Eq.~\eqref{eq:heff-matrix-elements} with the following coefficients on
the diagonal (details of the calculation are given in
Appendix~\ref{sec:effect-hamilt-example}):
\begin{widetext}
\begin{equation}
  \label{eq:e}
  \begin{split}
    e_1 & = \frac{1}{2} \left( - \frac{1}{1 + C_2} - \frac{1}{C_3} - \frac{1}{1
        + C_1 + C_3} - \frac{1}{C_2 + C_3} + \frac{1}{1 - C_1 - C_2 - C_3} -
      \frac{1}{1 + C_1} \right), \\ e_2  & = \frac{1}{2} \left( - \frac{1}{1 +
        C_2} - \frac{1}{C_3} + \frac{1}{1 - C_1 - C_3} - \frac{1}{C_2 + C_3} -
      \frac{1}{1 + C_1 + C_2 + C_3} - \frac{1}{1 + C_1} \right), \\ e_3 & =
    \frac{1}{2} \left( - \frac{1}{1 + C_2} - \frac{1}{C_3} - \frac{1}{1 + C_1 +
        C_3} - \frac{1}{C_2 + C_3} - \frac{1}{1 + C_1 + C_2 + C_3} + \frac{1}{1
        - C_1} \right),
  \end{split}
\end{equation}
and on the off-diagonal:
  \begin{equation}
  \label{eq:g1323}
  \begin{split}
    g_{13} & = \frac{1}{2^2} \left( - \frac{\frac{1}{C_2 + C_3} + \frac{1}{1 +
          C_1}}{1 + C_1 + C_2 + C_3} - \frac{\frac{1}{1 - C_1 - C_2 - C_3} -
        \frac{1}{C_2 + C_3}}{1 - C_1} + \frac{\frac{1}{1 - C_1 - C_2 - C_3} -
        \frac{1}{1 + C_1}}{C_2 + C_3} \right), \\ g_{23} & = \frac{1}{2^2}
    \left( -\frac{\frac{1}{1 + C_1} + \frac{1}{C_3}}{1 + C_1 + C_3} +
      \frac{\frac{1}{1 - C_1 - C_3} - \frac{1}{1 + C_1}}{C_3} - \frac{\frac{1}{1
          - C_1 - C_3} - \frac{1}{C_3}}{1 - C_1} \right),
  \end{split}
\end{equation}
and
\begin{multline}
  \label{eq:g12}  
  g_{12} = \frac{1}{2^3} \left( \frac{-\frac{\frac{1}{C_2 + C_3} + \frac{1}{1 +
          C_1 + C_3}}{1 + C_1 + C_2} + \frac{\frac{1}{1 - C_1 - C_2 - C_3} -
        \frac{1}{1 + C_1 + C_3}}{C_2} - \frac{\frac{1}{1 - C_1 - C_2 - C_3} -
        \frac{1}{C_2 + C_3}}{1 - C_1}}{C_3} \right. \\ - \frac{\frac{-
      \frac{1}{C_2 + C_3} - \frac{1}{C_3}}{C_2} - \frac{\frac{1}{1 - C_1 - C_2 -
        C_3} - \frac{1}{C_2 + C_3}}{1 - C_1} - \frac{\frac{1}{1 - C_1 - C_2 -
        C_3} - \frac{1}{C_3}}{1 - C_1 - C_2}}{1 - C_1 - C_3} \\ +
  \frac{-\frac{\frac{1}{1 + C_1 + C_3} + \frac{1}{C_3}}{1 + C_1} +
    \frac{\frac{1}{1 - C_1 - C_2 - C_3} - \frac{1}{1 + C_1 + C_3}}{C_2} -
    \frac{\frac{1}{1 - C_1 - C_2 - C_3} - \frac{1}{C_3}}{1 - C_1 - C_2}}{C_2 +
    C_3} \\ \left. + \frac{-\frac{\frac{1}{1 + C_1 + C_3} + \frac{1}{C_3}}{1 +
        C_1} - \frac{\frac{1}{C_2 + C_3} + \frac{1}{C_3}}{C_2} -
      \frac{\frac{1}{C_2 + C_3} + \frac{1}{1 + C_1 + C_3}}{1 + C_1 + C_2}}{1 +
      C_1 + C_2 + C_3} \right).
\end{multline}
\end{widetext}

The time scale $t_d$ can now be estimated as described in
Sec.~\ref{sec:estimation-t_d}. Evidently, also $t_d$ depends on the parameters
$C_p$, which we have to take into account when we minimize the cost
function~\eqref{eq:cost_function} for a given set of probabilities $p_n$,
$n = 1, 2 ,3$. For equal probabilities $p_n = 1/M$, we thus obtain $C_1 = 5.73$,
$C_2 = 0.19$, and $C_3 = 6.07$.  Optimizing these values iteratively we find
$C_1 = 7.91$, $C_2 = 0.24$, and $C_3 = 8.78$.  In Fig.~\ref{fig:results}~(c), we
show the exact time evolution of the squared amplitudes of the lowest three
instantaneous eigenstates. We note that in the lattice-gauge representation of
this example, the degenerate ground states have Hamming distances
$h_{13} = h_{23} = 3$ and $h_{12} = 4$.  Thus, as discussed below
Eq.~\eqref{eq:heff-matrix-elements}, the instantaneous ground states
$\ket{\varphi_n(t)}$ approach the states $\ket{x_n}$ for $t \to T$, and it is
indeed the probabilities of these states at $t = T$ we want to control. The
optimized $C_p$ lead to final amplitudes
$\abs{a_1}^2 = 0.344, \abs{a_2}^2 = 0.347,$ and $\abs{a_3}^2 = 0.309$, close to
their target values. We expect further improvement by including contributions of
higher order in the effective Hamiltonian. For this model, the optimization of
the $C_p$ can be carried out exactly, i.e., using the iterative procedure
described above with the \emph{full} quantum dynamics. Then we obtain
$C_1 = 9.31$, $C_2 = 0.40$, and $C_3 = 9.82$ with final probabilities almost
exactly $1/M$ [Fig.~\ref{fig:results}~(c) solid line]. This demonstrates that
the fidelity of the final state is not limited fundamentally but only by the
accuracy with which we determine the control parameters. For the second example
with probabilities shown in Fig.~\ref{fig:results}~(d), we obtain $C_1 = 5.53$,
$C_2 = 0.86$, and $C_3 = 2.44$, and by iterative optimization $C_1= 5.80$,
$C_2= 1.25$, and $C_3=2.68$, leading to
$\abs{a_1}^2 = 0.219, \abs{a_2}^2 = 0.297,$ and $\abs{a_3}^2 = 0.484$. The exact
results are again almost identical to the approximate solution. Let us stress
two features of the quantum dynamics: (i) The dynamics within the AMF is
evidently adiabatic, i.e., the occupation of the instantaneous ground state is
constant and close to unity, up to a time $t_d \approx 0.5$ and
$t_d \approx 0.6$ in Figs.~\ref{fig:results}~(c) and~\ref{fig:results}~(d), respectively. (ii) This
time $t_d$ agrees well with the time at which the cost function, evaluated for
the instantaneous ground state, takes its minimum --- in line with the above
claim that the desired state should be reached already at $t_d$ and, in
particular, prior to $T$.

\subsection{Robustness to constraint errors}
\label{sec:robustn-constr-error}

\begin{figure}
  \centering
  \includegraphics[width=6.5cm]{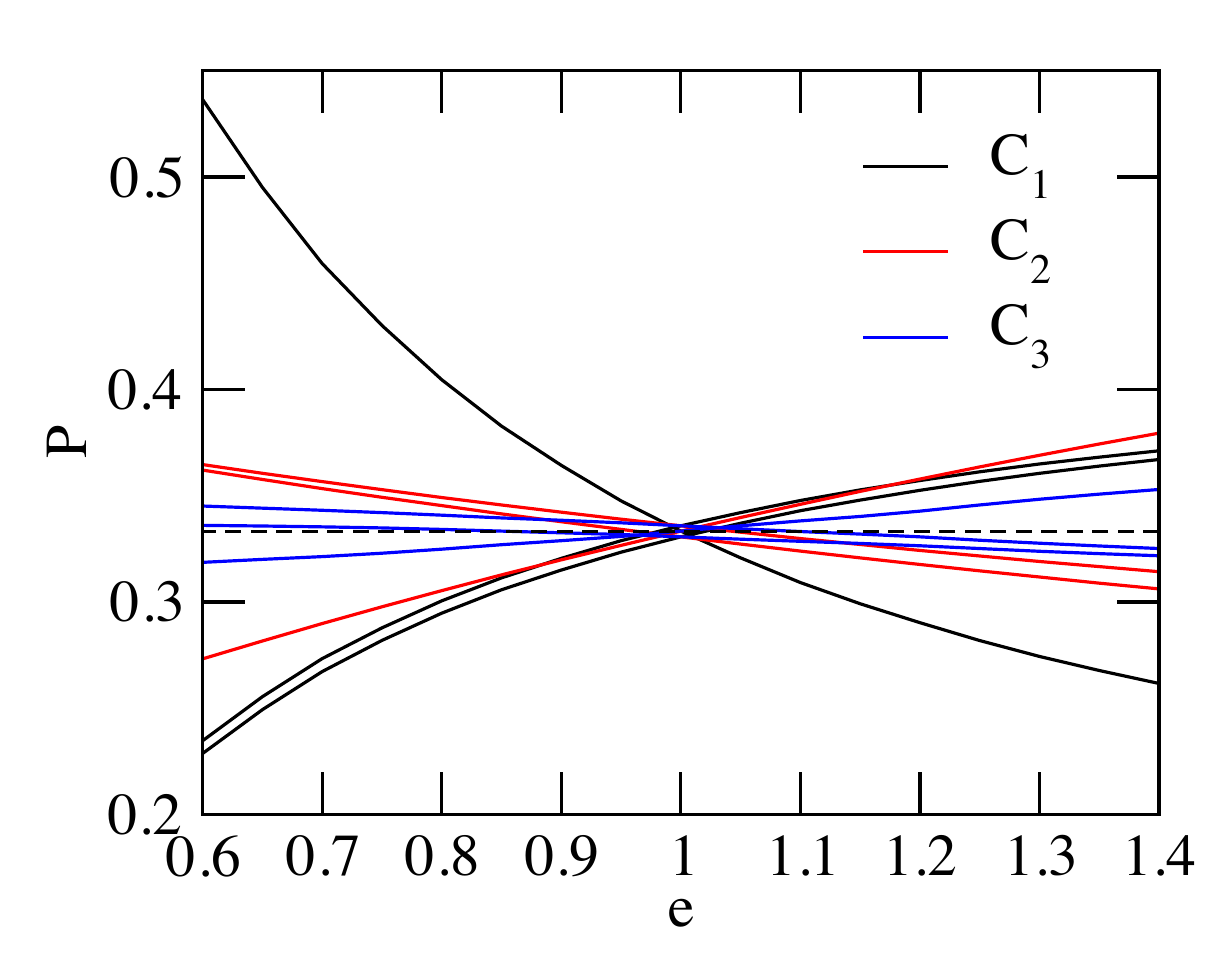}
  \caption{Probabilities $P$ of the final superposition as a function of the
    relative error in the constraints $e$. We consider the example shown in
    Fig.~\ref{fig:fig1} with equal target probabilities including errors in
    $C_1$ (black), $C_2$ (red) and $C_3$ (blue). The three lines of each color
    show the probabilities $\abs{a_1}^2, \abs{a_2}^2,$ and $\abs{a_3}^2$,
    respectively. For the constraint $C_1$ which is most sensitive to errors,
    the deviation in $P$ is $10\,\%$ for a error of $20\,\%$ in $C_1$. For other
    constraints the deviation in $P$ is even smaller.}
  \label{fig:stability}
\end{figure}

As we demonstrate systematically in the following, the final probabilities are
rather robust to with respect to errors in the control parameters $C_p$. We
define the relative error $e$, where $\hat{C_p} = e C_p$, and $C_p$ are the
constraints obtained by optimization as explained above. For the example given
in Fig.~\ref{fig:fig1} and target probabilities $p_n = 1/M$, the values for
$C_p$ are taken from the exact calculation with the full quantum dynamics. The
changes in the probabilities for $e$ ranging from $0.6$ to $1.4$ are shown in
Fig.~\ref{fig:stability}. Around $e=1$, the solution is remarkably insensitive
to errors.

\subsection{Ergodicity of the solution space}
\label{sec:ergod-solut-space}

Finally, to address how general the method is, we ask whether it is possible to
reach every combination of $\abs{a_n}^2$ from the available parameters $C_p$ and
$t_d$. Let us count the number of degrees of freedom first: The number of
constraints is $K-N+1$ which grows quadratically with the length of the bit
strings $N$, while the number of variables $\abs{a_n}^2$ is $M-1$ because of the
normalization condition $\sum_{n = 1}^M p_n =1$. Thus, $M - 1$ degrees of
freedom need to be programed and $K$ degrees of freedom are available. This
means that for large systems the number of available parameters will in general
be large as compared to the number of variables that we want to program, leading
us to expect that it should be possible to prepare any superposition.

\begin{figure*}
  \centering
  \includegraphics[width=16cm]{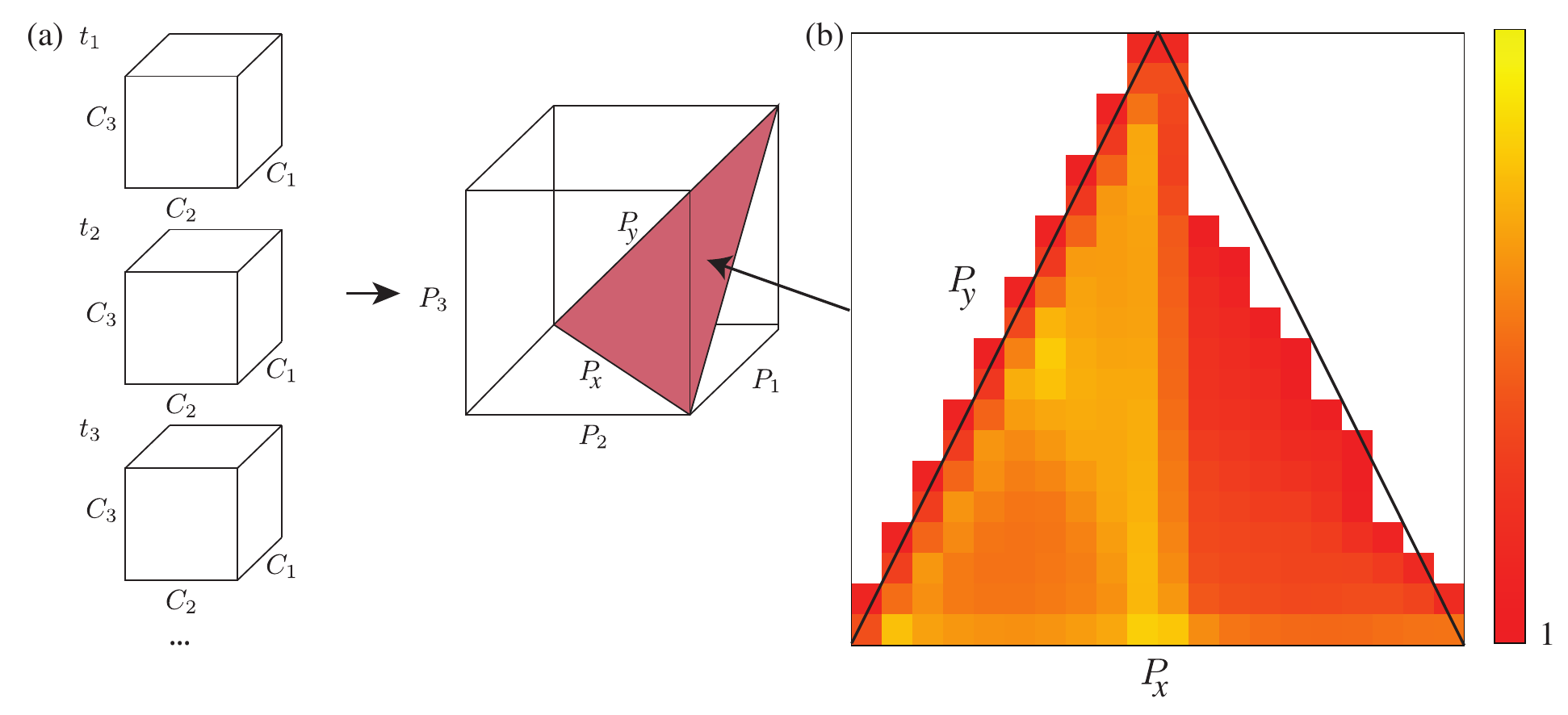}
  \caption{(a) Our protocol maps the parameter set $\{t_d, C_1,C_2,C_3\}$ to a
    two-dimensional plane in the solution space spanned by $\{p_1, p_2, p_3\}$.
    The mapping is ergodic, if each point on this plane can be reached. (b)
    Heatmap of the projected probabilities on the two-dimensional solution
    plane. Points that can be reached (red) are all within the expected
    triangle, while a small region in the top right (white) cannot be reached.}
  \label{fig:ergodicity}
\end{figure*}

For small systems, we can systematically address the question whether the
solution hyperplane is ergodically populated from all constraint
combinations. In the example introduced above, the number of bit strings is
$M=3$. With the condition $p_1 + p_2 + p_3 = 1$, the solution space is a
two-dimensional plane in the three-dimensional space of probabilities. We can
systematically check how this two-dimensional plane is populated from the
following simplified model, illustrated in
Fig.~\ref{fig:ergodicity}~(a). According to the effective model, we assume that
the amplitudes in the instantaneous ground state stay constant after the
freezing time $t_d$. We further assume that the freeze-in can be at any time
during the protocol. Therefore, we can check each combination of $C_1$, $C_2$,
and $C_3$ for each time $t_d$. We scanned the constraints in the interval $0.1$
to $4.0$ in steps of $0.1$ and the time in steps of $T/30$. The resulting
solution plane is shown in Fig.~\ref{fig:ergodicity}~(b). Even in this
simplified model, the mapping is almost ergodic. For larger systems we expect
full ergodicity due to the larger number of degrees of freedom.

\section{Conclusions}
\label{sec:conclusions}

We presented a general framework that utilizes lattice-gauge degrees of freedom
to control the dynamical preparation of a quantum superposition with tunable
weights. The control parameters are determined from an effective multi-level
Landau-Zener model. Our protocol can be implemented in state-of-the-art
experiments, e.g., neutral atoms~\cite{lhzrydberg} or superconducting
qubits~\cite{Puri2017}. From an application point of view, we hope this work is
useful for quantum algorithms that benefit from data provided as
superpositions~\cite{qram}. The effective model also suggests a general answer
to the question for the conditions of fair
sampling~\cite{matsuda2009ground,KatzgraberFair}. Our protocol opens several
routes for extension, e.g., in combination with shortcut-to-adiabaticity
methods~\cite{berry2009transitionless,deffner2014classical,Sels16052017,patra2017shortcuts},
optimal control~\cite{optimalcontrol} and quantum approximate optimization
algorithms~\cite{farhi2014quantum}. An interesting challenge for future research
is controlling not only the weights but also the phases in the final
superposition state~\cite{Jordan2012}.

\begin{acknowledgments}
  We thank H. Katzgraber, S. Mandra, P. Zoller, P. Hauke, H. Neven, N. Ding,
  M. Mohseni, S. Gazit, M. Serbyn, and E. Altman for fruitful
  discussions. Research was funded by the Austrian Science Fund (FWF) through a
  START grant under Project No. Y1067-N27, the Hauser-Raspe foundation, and by
  the ERC through the synergy grant UQUAM.
\end{acknowledgments}

\appendix

\section{Effective Hamiltonian for the example in Sec.~\ref{sec:example}}
\label{sec:effect-hamilt-example}

Using Eq.~\eqref{eq:heff}, it is straightforward to calculate $\heff(t)$
numerically. However, for the simple example considered in
Sec.~\ref{sec:example}, we find it instructive to obtain explicit expressions
for the matrix elements of $\heff(t)$ by hand. This also clarifies how the
constraint strengths $C_p$ enter $\heff(t)$ and thus allow us to control the
quantum dynamics.

For the model introduced in Fig.~\ref{fig:fig1} of the main text, the part of
the Hamiltonian involving the constraints involves two three-body and one
four-body interaction,
\begin{gather}
  \label{eq:HC}
  H_C = - C_1 \sz_1 \sz_2 \sz_4 - C_2 \sz_2 \sz_3 \sz_5 - C_3 \sz_2 \sz_4 \sz_5
  \sz_6.
\end{gather}
Equivalently, the three-body terms can be viewed as four-body terms in which one
spin is held fixed, e.g., by an external field~\cite{lechner2015}. We denote by
$S_i$ the set of constraints that involve spin $i$,
\begin{equation}
  \label{eq:Si}
  \begin{gathered}
    S_1 = \{ 1 \}, \quad S_2 = \{ 1, 2, 3 \}, \quad S_3 = \{ 2 \}, \\ S_4 = \{
    1, 3 \}, \quad S_5 = \{ 2, 3 \}, \quad S_6 = \{ 3 \}.
  \end{gathered}
\end{equation}
The energy of the degenerate ground states $\ket{z_n}$ is
\begin{equation}
  \label{eq:E0}
  E_0 = - \frac{t}{T} \left( \sum_i J_i z_{n, i} + \sum_{p = 1}^3 C_p \right),
\end{equation}
where the contribution due to the local fields depends on the orientation of the
spins $z_{n, i}$ (here, we replace $z_{n, i} = 0,1 \to \pm 1$), while each
constraint contributes with $- t C_p/T$ to lowering the energy of states in the
ground-state manifold.

We obtain the diagonal elements of $\heff$ at second-order in perturbation
theory:
\begin{equation}
  \label{eq:Heffnn1}
  \begin{split}
    H_{\mathrm{eff}, nn} & = E_0 + \left( 1 - \frac{t}{T} \right)^2 \sum_{i, j}
    \bra{z_n} \sx_i \frac{1}{E_0 - H_0} \sx_j \ket{z_n} \\ & = E_0 + \left( 1 -
      \frac{t}{T} \right)^2 \sum_i \frac{1}{E_0 - \bra{z_n} \sx_i H_0 \sx_i
      \ket{z_n}}.
  \end{split}
\end{equation}
Here we used that the state $\sx_i \ket{z_n}$ is still a product state in the
$\sz$-basis (only with spin $i$ flipped relative to $\ket{z_n}$), and $H_0$ is
diagonal in this basis. The excitation energy of the state $\sx_i \ket{z_n}$ is
given by
\begin{equation}
  \bra{z_n} \sx_i H_0 \sx_i \ket{z_n} - E_0 = \frac{2 t}{T} \left(
    J_i z_{n, i} + \sum_{p \in S_i} C_p \right),
\end{equation}
i.e., we get a contribution from the local field acting on spin $i$ and from all
the constraints that involve this same spin. These constraints are satisfied in
the state $\ket{z_n}$ and therefore violated if the single spin $i$ is
flipped. Hence,
\begin{equation}
  \label{eq:Heff_d}
  H_{\mathrm{eff}, nn} = E_0 - \frac{T}{2 t} \left( 1 - \frac{t}{T} \right)^2 \sum_i
  \frac{1}{J_i z_{n, i} + \sum_{p \in S_i} C_p}.
\end{equation}

For the off-diagonal elements $H_{\mathrm{eff}, nn'}$ there is no equally
compact form. However, they can be visualized in an intuitive way. The key
observation is that a given order of the perturbation, applied to a state
$\ket{z_{n'}}$ leads to ``paths'' originating from this state. The length of
these paths is the number of single spin flips to go from one state to the
other, and the required order of perturbation theory is determined by the length
of paths connecting two states $\ket{z_{n'}}$ and $\ket{z_n}$: To leading order,
those paths contribute to $H_{\mathrm{eff}, nn'}$ that connect the states
$\ket{z_{n'}}$ and $\ket{z_n}$ in the minimum number of steps. This number is
given by the Hamming distance $h_{nn'}$ of these states.

In our example, two of the degenerate ground states of $H_0$ are given by
$\ket{z_n} = \ket{010011}$ and $\ket{z_{n'}} = \ket{001001}$, i.e., they have
Hamming distance $h_{nn'} = 3$. In particular, $\ket{z_n}$ can be obtained from
$\ket{z_{n'}}$ by flipping the spins at positions 2, 3, and 5,
$\ket{z_n} = \sx_2 \sx_3 \sx_5 \ket{z_{n'}}$. Then, there is an off-diagonal
matrix element $H_{\mathrm{eff},nn'}$ in third order in perturbation theory. As
explained above, the terms contributing to this matrix element can be visualized
as sums over paths connecting the states $\ket{z_n}$ and $\ket{z_{n'}}$. Each
path corresponds to a particular order in which the spins 2, 3, and 5 are
flipped, and each segment of a given path contributes with a factor of minus one
over the excitation energy of the state at the end of the segment,
\\[.2cm]
\includegraphics[width=\linewidth]{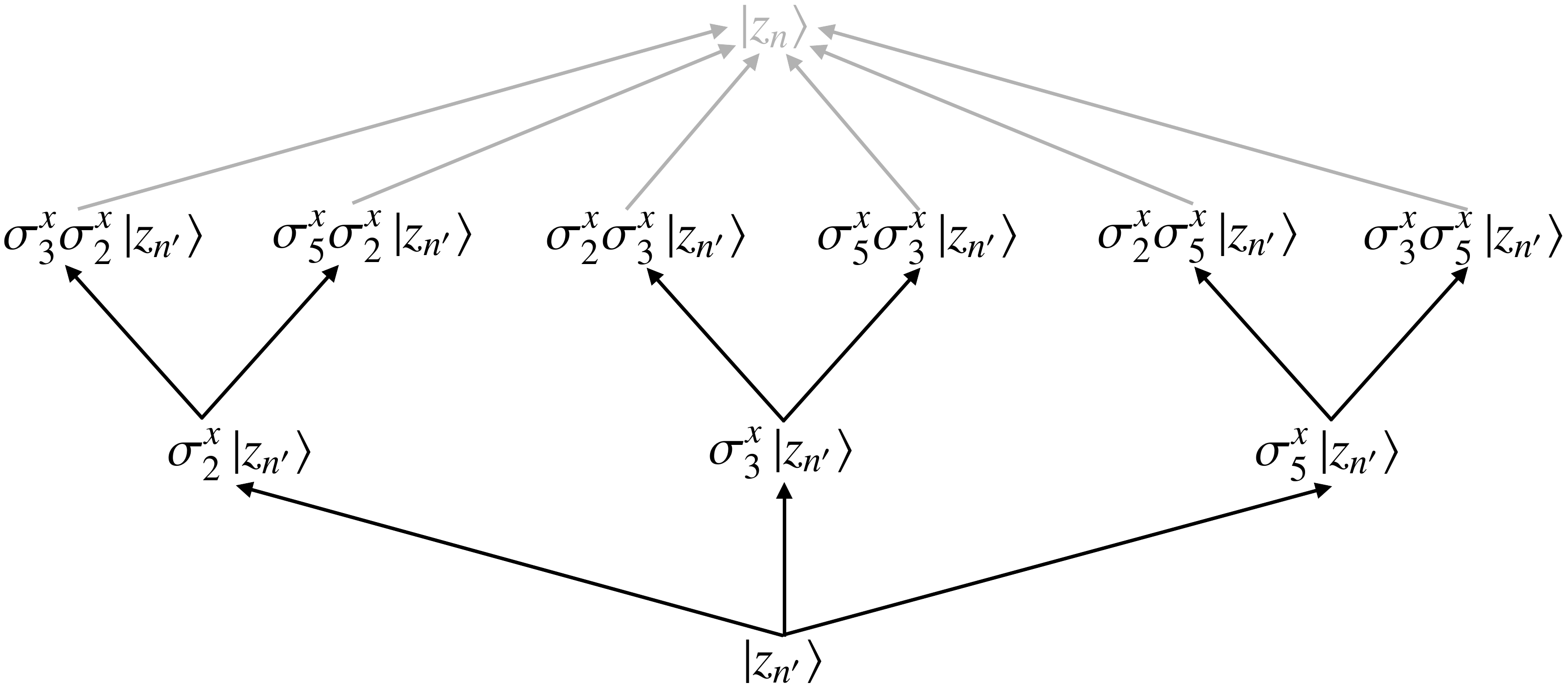}
\\[.2cm]
The tree-like structure of the diagram translates directly to the iterated
fractions, e.g., in the expression for $g_{13}$ in Eq.~\eqref{eq:g1323}.
Indeed, comparing the first term in $g_{13}$ and the left-most branch of the
tree, we see that flipping spin 2 leads to a state with excitation energy
$2 \left( 1 + C_1 + C_2 + C_3 \right)$; depending on which spin is flipped next,
states with energies of $2 \left( 1 + C_1 \right)$ or
$2 \left( C_2 + C_3 \right)$ are reached. The last segment of the path that
leads to $\ket{z_n}$ does not contribute to the matrix element
$H_{\mathrm{eff},nn'}$ as is indicated in the diagram by reduced opacity.

We formalize these considerations in the following. Let's denote the spins we
have to flip to get from $\ket{z_{n'}}$ to $\ket{z_n}$ by
$D = \left( i_1, i_2, \dotsc, i_h \right)$ with $h = h_{nn'}$. The paths that
contribute to $H_{\mathrm{eff}, nn'}$ are different sequences of spin flips,
i.e., permutations of $D$. We denote such a permutation by
$\pi(D) = \left( i_{\pi_1}, \dotsc, i_{\pi_h} \right)$. There are $h!$
permutations. For a given permutation, the repeated action of the perturbation
$V$ takes us through the sequence of states
\begin{multline}  
  \ket{z_{n'}} \to \sx_{i_{\pi_1}} \ket{z_{n'}} \to \sx_{i_{\pi_2}}
  \sx_{i_{\pi_1}} \ket{z_{n'}} \\ \to \dotsb \to \sx_{i_{\pi_h}} \dotsb
  \sx_{i_{\pi_1}} \ket{z_{n'}} = \ket{z_n}.
\end{multline}
Using the same notation as in Eq.~\eqref{eq:Heffnn1}, we find
\begin{multline}
  \label{eq:Heff_od}
  H_{\mathrm{eff}, nn'} = \left( -1 \right)^h \left( 1 - \frac{t}{T} \right)^h
  \\ \times \sum_{\pi} \frac{1}{E_0 - \bra{z_{n'}} \sx_{i_{\pi_1}} \dotsb \sx_{i_{\pi_{h -
          1}}} H_0 \sx_{i_{{\pi_{h - 1}}}} \dotsb \sx_{i_{\pi_1}} \ket{z_{n'}}}
  \\ \times \dotsb \frac{1}{E_0 - \bra{z_{n'}} \sx_{i_{\pi_1}} H_0
    \sx_{i_{\pi_1}} \ket{z_{n'}}}.
\end{multline}
The constraint strengths $C_p$ enter this expression through the energy
denominators as above in Eq.~\eqref{eq:Heff_d}. While Eqs.~\eqref{eq:Heff_d}
and~\eqref{eq:Heff_od} provide valuable insight into the general structure of
the effective model, for doing any practical calculation it is much simpler to
directly implement Eq.~\eqref{eq:heff}, e.g., in \textsc{Mathematica}. Doing
this for our model, we find the explicit expressions for the coefficients $e_n$
and $g_{nn'}$ reported in Eqs.~\eqref{eq:e},~\eqref{eq:g1323},
and~\eqref{eq:g12}.

\bibliography{references.bib}

\begin{thebibliography}{45}
\bibitem [{tob()}]{tobepublished}%
  \BibitemOpen
  \href@noop {} {\bibinfo  {journal} {To be published}\ }\BibitemShut {NoStop}%
\end{thebibliography}

\end{document}